\documentclass[onecolumn,aps,nofootinbib]{revtex4}
\usepackage{amssymb, amsmath, bm, dcolumn, epsf, graphicx, latexsym, mathbbol, slashed}
\usepackage{graphicx}
\usepackage{dcolumn}
\usepackage{bm}
\usepackage{times}
\usepackage{amssymb}
\usepackage{color}
\usepackage{multirow}
\usepackage[colorlinks,linkcolor=blue,citecolor=blue,urlcolor=blue]{hyperref}

 \def\tskip{\setlength{\tskip}{5pt}}
\def\colwidth{\setlength{\colwidth}{3.5in}}

\newcommand{\lsim}{\mathrel{\hbox{\rlap{\lower.55ex\hbox{$\sim$}} \kern-.3em \raise.4ex \hbox{$<$}}}}
\newcommand{\gsim}{\mathrel{\hbox{\rlap{\lower.55ex\hbox{$\sim$}} \kern-.3em \raise.4ex \hbox{$>$}}}}
\newcommand{\beq}{\begin{equation}}
\newcommand{\eeq}{\end{equation}}
\newcommand{\be}{\begin{equation}}
\newcommand{\ee}{\end{equation}}
\newcommand{\bes}{\begin{equation*}}
\newcommand{\ees}{\end{equation*}}
\newcommand{\beqa}{\begin{eqnarray}}
\newcommand{\eeqa}{\end{eqnarray}}
\newcommand{\bea}{\begin{eqnarray}}
\newcommand{\ena}{\end{eqnarray}}

\usepackage{color}

\usepackage{ulem}

\begin{document}

\title{Constraining the time variation of Newton's constant $G$ with gravitational-wave standard sirens and supernovae}

\author{Wen Zhao$^{1,2}$, Bill S. Wright$^{3}$, Baojiu Li$^{4}$}
\affiliation{$^{1}$ CAS Key Laboratory for Researches in Galaxies and Cosmology, Department of Astronomy, University of Science and Technology of China, Chinese Academy of Sciences, Hefei, Anhui 230026, China \\
$^{2}$ School of Astronomy and Space Science, University of Science and Technology of China, Hefei 230026, China \\
$^{3}$ Institute of Cosmology \& Gravitation, University of Portsmouth, Portsmouth, Hampshire, PO1 3FX, UK \\
$^{4}$Institute for Computational Cosmology, Department of Physics, Durham University, Durham DH1 3LE, UK}


\begin{abstract}
The intrinsic peak luminosity of Type Ia supernovae (SNIa) depends on the value of Newton's gravitational constant $G$, through the Chandrasekhar mass $M_{\rm Ch}\propto G^{-3/2}$. If the luminosity distance can be independently determined, the SNIa can be treated as a tracker to constrain the possible time variation of $G$ in different redshift ranges. The gravitational-wave (GW) standard sirens, caused by the coalescence of binary neutron stars, provide a model-independent way to measure the distance of GW events, which can be used to determine the luminosity distances of SNIa by interpolation, provided the GW and SNIa samples have similar redshift ranges. We demonstrate that combining the GW observations of third-generation detectors with SNIa
data provides a powerful and model-independent way to measure $G$ in a wide redshift range, which can constrain the ratio $G/G_0$, where $G$ and $G_0$ are respectively the values in the redshift ranges $z>0.1$ and $z<0.1$, at the level of $1.5\%$.

\end{abstract}


\maketitle

\section{Introduction}

The measurement of Newton's constant $G$ is one of the key tasks in modern physics. In General Relativity, $G$ is assumed to be constant. However, in general alternative theories of gravity, it can become both time- and space-dependent. { {For instance, in Brans-Dicke gravity, the value of $G$ is inversely proportional to the mean value of the scalar field $\phi$ in the Universe, which evolves with the expansion of the Universe \cite{Weinberg2}. While in general screened modified gravity, which is a kind of scalar-tensor theory with screening mechanisms, the value of $G$ depends both on the mean value of the scalar field $\phi$ in the Universe and on the local Newtonian potential of the observed object \cite{zhang2016}.}} Numerous methods have been proposed to measure $G$ on different time scales, including the lunar ranging experiment \cite{lunar}, pulsar timing observations \cite{pulsar}, Big Bang nucleosynthesis (BBN) observations \cite{BBN} and so on \cite{G-review,G-review2}.

Type Ia supernovae (SNIa) are `standard candles' in the standard cosmological model \cite{SNIa-review}. However, analytical models of their light curves generally predict that the absolute magnitude of a SNIa depends on the value of
$G$. Therefore, measuring the absolute magnitude of
SNIa can determine the values of $G$ at different redshifts. To
achieve this, the luminosity distance $d_{\rm L}$ to each SNIa should be independently determined, which is requisite to fix the absolute magnitude of
SNIa from observations. In previous works \cite{SNIa-G-previous-work1,SNIa-G-previous-work2}, the independent determination of $d_{\rm L}$ is given by assuming specific cosmological models. {{For instance, in \cite{SNIa-G-previous-work1} the authors determined the distance $d_{\rm L}$ of SNIa by assuming a flat universe without cosmological constant but with a varying $G$ as a function of redshift. While in \cite{SNIa-G-previous-work2}, the authors assumed the $\Lambda$CDM model, or a polynomial form of the Hubble parameter. These assumptions induce that the resulting constraints on $G$ in these papers are model-dependent.}} We avoid this issue by considering the potential observations of GW standard sirens in
similar redshift ranges to those of the SNIa, which provide the desired independent measurement of $d_{\rm L}$ { \footnote{In addition to the GW standard sirens, the Cepheid variables can also be used to measure the distance $d_{\rm L}$ of SNIa, if we assume their empirical period-luminosity relation \cite{Phillipse}. However, this measurement suffers from two defects: First, only the nearby Cepheid variables are observable, so this method is applicable only for extremely low-redshift range. Second, this method depends on the so-called cosmic distance ladder \cite{Weinberg}. }}. Therefore, combining SNIa and GW data provides a novel way to measure $G$ on a cosmological scale. In this method, the constraint
on $G$ is at the time of the SNIa. Thus, once a sufficient number of events have been observed, a constraint map, as a function of redshift, could be constructed.

\section{Gravitational dependence of SNIa}

The empirical observation of SNIa gives the distance estimation, under the assumption that all SNIa have the same intrinsic luminosity if they are identical in colour, shape and galactic environment \cite{1401.4064}. This model expresses the distance modulus in an isotropic universe, $\mu=5\log_{10}(d_{\rm L}/{\rm 10pc})$, as
\begin{equation}\label{eq11}
\mu=m_{\rm B}^*-(M_{\rm B}-\alpha\times X_1+\beta\times {C}),
\end{equation}
where $m_{\rm B}^*$ is the observed peak magnitude in the rest-frame $B$ band, $X_1$ is a time stretching of the light-curve, and $C$ is a supernova colour at maximum brightness. For any given SNIa, these quantities are obtained from a fit to the light-curve model of SNIa. $\alpha$ and $\beta$ are the nuisance parameters. The absolute magnitude $M_{\rm B}$ depends on the host galaxy properties, which can be approximately corrected by assuming that $M_{\rm B}$ is related to the host stellar mass through $M_B=M^1_B$ if the host stellar mass is smaller than $10^{10}$ solar masses while $M_B=M^1_B+\Delta_{M}$ if otherwise, where $\Delta_{M}=-0.08$mag. The absolute magnitude $M^1_B$ is calibrated for $X_1=C=0$ and is treated as a constant \cite{1401.4064}, relating to the calibrated intrinsic luminosity $L$ as $M_B^{1}=M_{\odot}-2.5\log_{10}(L/L_{\odot})$, where $M_{\odot}$ and $L_{\odot}$ are the Sun's absolute magnitude and luminosity respectively.

\begin{figure}
\begin{center}
\centerline{\includegraphics[width=15cm]{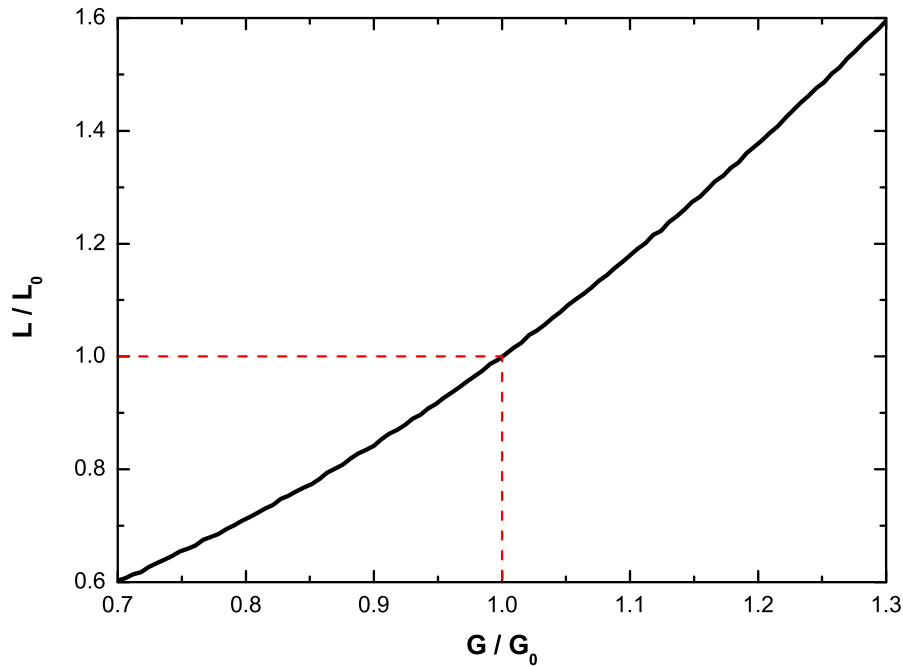}}
\end{center}\caption{The {calibrated} intrinsic luminosity $L$ of {SNIa} as a function of $G$, where $G_0$ is the present-day value, and $L_0$ is the absolute luminosity of {SNIa} at $G=G_0$.} \label{f2}
\end{figure}

In theories of modified gravity with variable $G$, the $M_{\rm B}^{1}$ of
SNIa is not a constant. Analytical models of the light curve predict that the peak luminosity of a SNIa is proportional to the mass of nickel synthesised \cite{Arnett1982}, which is approximated to be proportional to the
Chandrasekhar mass $M_{\rm Ch}$ \cite{Gaztanaga2002}. {{$M_{\rm Ch}$ is the theoretical limit around which an accreting white dwarf will undergo supernova \cite{Chandrasekhar}.}} Since $M_{\rm Ch}$ depends on $G$, the measurement of peak luminosity of
SNIa can be used to determine the variation of $G$ with redshift. In \cite{1710.07018},
the full effect of a time variation of $G$ on the combined UV+optical+IR SNIa light curves was investigated by a semi-analytic analysis. This method enabled calibration of luminosities based on time-stretching.
In this treatment, the main effect is a change of the Chandrasekhar mass, $M_{\rm Ch}\propto G^{-3/2}$. If $G$ deviates from the its present-day value $G_0$, $M_{\rm Ch}$ differs from $1.44$ solar masses, which
leads to the modification of the time-stretch calibrated intrinsic luminosity $L$. {{The physics behind why changing $M_{\rm Ch}$ affects the time-stretch calibrated intrinsic luminosity is as follows. A larger $M_{\rm Ch}$ will lead to a larger mass of ejecta in the aftermath of the supernova explosion, which in turn will hinder the transmission of radiation through the ejecta. This results in a fainter, wider light curve. When this light-curve is rescaled to calibrate for the time-stretching, the decrease in width requires a further matching decrease in luminosity. Thus there is a negative relation between $M_{\rm Ch}$ and the time-stretch calibrated intrinsic luminosity $L$, and since $M_{\rm Ch}\propto G^{-3/2}$, a positive relationship between $G$ and $L$. A full discussion of this effect is contained in \cite{1710.07018}.}
The predicted $L$ as a function of $G$ using the analysis of \cite{1710.07018} is presented in Fig.~\ref{f2}, which shows
a sensitive dependence of $L$ on $G$.} Therefore, if the value of $L$, or equivalently the $M^1_{B}$, can be determined at different redshifts, we could infer the local value of $G$, which provides a novel method to measure
$G$ in different redshift ranges. { {Note that in this work we have assumed that the time variation of $G$ dominates, while its spatial variation is negligible, and we have not considered possible screening effects commonly encountered in modified gravity models. For screened models, the variation of $G$ can be smaller in dense environments such as galaxies, and this will make the time-dependence of the luminosity weaker \cite{zhang2016}.}}

\section{GW standard sirens}

From Eq.~(\ref{eq11}), we observe that for any given {SNIa}, the value of $M_{\rm B}^1$ can be derived if one can independently measure the luminosity distance $d_{\rm L}$. GW standard sirens provide a model-independent way to
achieve this. From the observations of GW signals, caused by coalescence of binary neutron stars (BNSs), one can obtain the $d_{\rm L}$ of a GW event in an absolute way, without having to rely on a cosmic distance ladder \cite{schutz}. In many cases, it is also possible to identify their electromagnetic counterparts and determine their redshifts \cite{em-counterparts,em-counterpart2,Dave}.
Therefore, this provides a novel way to construct the Hubble diagram over a wide redshift range. The third-generation (3G) GW experiments can detect the high-redshift GW signals. By combining the $d_{\rm L}$ and $z$ of standard sirens, we can directly construct the distance modulus as a function of $z$ for a wide redshift range.

Two proposals are currently under consideration for 3G GW detectors: the Einstein Telescope (ET) in Europe \cite{et}, and the Cosmic Explorer (CE) in the U.S.~\cite{ce}.
The coordinates and orientations of ET and CE are given in Table 1 of \cite{zhao2017}, and the amplitude spectral densities are given by Fig.~1 of \cite{zhao2017}. ET consists of three Michelson interferometers, and interarm angle of $60^{\circ}$, arranged to form an equilateral triangle, and we adopt the ET-D configuration in this paper \cite{et}. We consider a 3G network consisting of ET and CE,
and summarise the main results as follows. The response of an incoming GW signal is a linear combination of two wave polarisations, $d_I(t)=F_I^{+}h_+(t)+F_I^{\times}h_{\times}(t)$. The detector's beam-pattern functions $F_I^{+}$ and $F^{\times}_I$ depend on the source localization $(\theta_s,\phi_s)$ and the polarisation angle $\psi_s$. The restricted post-Newtonian approximated waveforms $h_+$ and $h_{\times}$ for the non-spinning BNSs depend on the mass ratio $\eta\equiv m_1m_2/(m_1+m_2)^2$, the chirp mass $\mathcal{M}_c\equiv(m_1+m_2)\eta^{3/5}$ ($m_1$ and $m_2$ are the physical masses of stars), the $d_{\rm L}$, the inclination angle $\iota$, the merging time $t_c$ and merging phase $\psi_c$ \cite{sathya}. So, for a given BNS, the response of detector depends on ($\mathcal{M}_c,\eta,t_c,\psi_c,\theta_s,\phi_s,\psi_s,\iota,d_{\rm L}$). Employing the nine-parameter Fisher matrix $\Gamma_{ij}$ and marginalising over the other parameters, we derive the uncertainty $(\Gamma^{-1})_{ii}^{~1/2}$ for the $i$-th parameter in analysis, with $i=9$ for $d_{\rm L}$ (see \cite{zhao2017} for details).

In the low-$z$ range, it is possible to identify the electromagnetic counterparts (e.g., kilonovae \cite{kilonovae}) of GW events and fix their redshifts. We numerically simulate the BNS samples with random binary orientations and sky directions. The redshifts are uniformly distributed in comoving volume in the range $z<0.1$. Current observation of the GW170817 burst predicts the event rate in the range of $[320,4740]{\rm Gpc}^{-3}{\rm year}^{-1}$\cite{GW170817PRL}. Assuming three-year observations by a 
3G network, we expect to observe $[3.0\times 10^{2},4.5\times 10^{3}]$ events at $z<0.1$, and $[5.4\times 10^{5},8.1\times 10^{6}]$ events at $z<2$. We randomly select 1000 samples with $z<0.1$ to mimic the detections of 3G network in low-$z$ range. In addition, a pessimistic case with $300$ events, and an optimistic case with $4500$ events are also discussed below for comparison. For each sample, distance measurements include two kinds of uncertainties: the instrumental error $\Delta d_{\rm L}$ calculated above, and an error $\tilde{\Delta} d_{\rm L}$ due to the effects of weak lensing, which can be assumed as $\tilde{\Delta} d_{\rm L}/d_{\rm L}=0.05z$ \cite{sathya2009}. Thus, the total uncertainty is $\sigma_{d_{\rm L}}=[(\Delta d_{\rm L})^2+(\tilde{\Delta} d_{\rm L})^2]^{1/2}$.

\begin{figure}
\begin{center}
\centerline{\includegraphics[width=15cm]{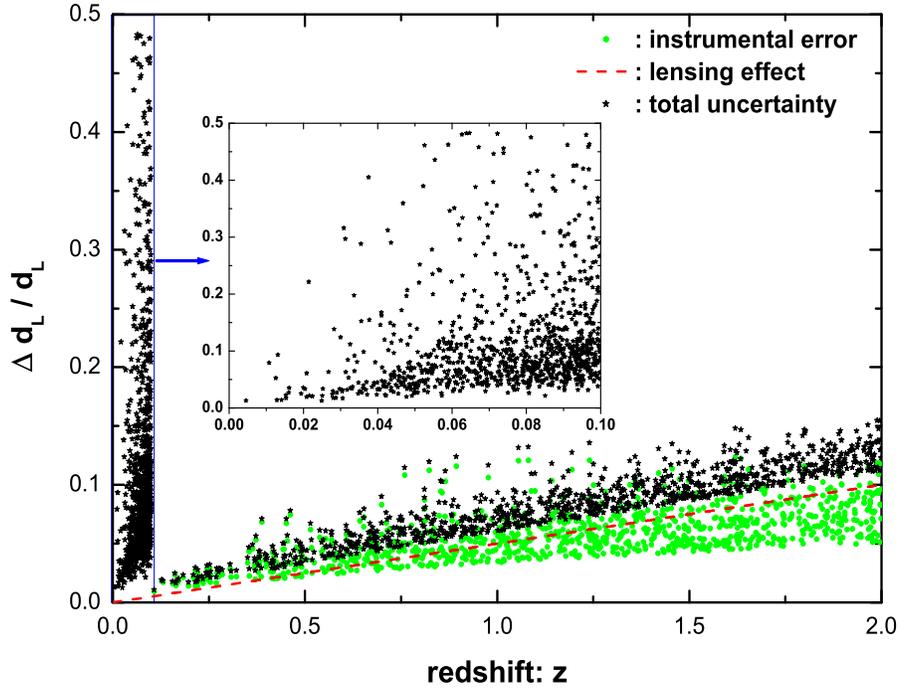}}
\end{center}\caption{The values of $\Delta d_{\rm L}/d_{\rm L}$ (green dot), $\tilde{\Delta} d_{\rm L}/d_{\rm L}$ (red line), $\sigma_{d_{\rm L}}/d_{\rm L}$ (black star) for the simulated GW samples, including 1000 BNSs at $z<0.1$, and 1000 face-on ones at $0.1<z<2$. Note that, for the low-$z$ events, we have not presented the results of $\Delta d_{\rm L}/d_{\rm L}$, which are overlapped with the corresponding results of $\sigma_{d_{\rm L}}/d_{\rm L}$.}\label{f1}
\end{figure}

For high-$z$ BNSs, the promising method to measure their redshifts is to observe their short-hard $\gamma$-ray burst (shGRB) counterparts. However, the $\gamma$ radiation is emitted in a narrow cone nearly perpendicular to the binary orbital plane, and the observed shGRBs are nearly all beamed towards the Earth \cite{shGRB}. For these face-on binaries, the parameters ($\theta_s,\phi_s,\iota,\psi_s$) can be fixed by electromagnetic observations. We repeat the calculation for a large number of face-on GW sources assuming a uniform distribution in comoving volume for a redshift range of $0.1<z<2$, as stated above, but adopting a five-parameter Fisher matrix. Although about $10^{6}$ GW events are expected to be observed, only a small partial of them can be treated as standard sirens with measured redshifts. Similar to previous works \cite{sathya2009,zhao2017}, we conservatively estimate that only 1000 BNSs are used as standard sirens, which are randomly chosen to mimic the observations in high-$z$ range. Fig.~\ref{f1} presents the redshift distribution and uncertainty of $d_{\rm L}$ for the samples in both low-$z$ and high-$z$ ranges. Note that, we also consider the case with 2000 events for comparison. Following \cite{hu}, for each BNS, we take the value of $d_{\rm L}$ to be the exact value of a given model, so we expect constraints to be centered on the fiducial parameter values rather than displaced by $\sim 1\sigma$. These constraints can be thought of as the average over many possible realisations of the data.

{{Note that in some specific modified gravity theories, for instance theories with time-dependent effective Planck mass \cite{Luca2017} or non-local modifications \cite{Belgacem2017}, the effective luminosity distance of GW could be different from that of electromagnetic waves. Comparison of these two distances provides a novel way to test these gravitational theories \cite{Luca2017,Belgacem2017}. Unlike such previous works, in this paper we consider a phenomenological theory of gravity, which is the same as GR but allowing the value of $G$ to be time-dependent. The waveforms of GW in this theory have been explicitly studied in the literature \cite{yunes}. Although the definition of $d_{\rm L}$ is the same as in GR, a time-dependent $G$ can generally revise the GW waveform of compact binaries, which could influence the determination of $d_{\rm L}$ from GW observations \cite{yunes}. However, for the GW events of BNS coalescence observed by 3G network, the duration
is typically several minutes, and the variation of $G$ during the GW burst is negligible \cite{yunes}. Therefore, for each GW event, we can consider $G$ to be
constant. For this case, GW waveforms depend on $G$ only through the combination of $Gm_1$ (or $Gm_2$) \cite{Maggiore}, i.e. the NS masses and $G$ value are completely degenerate, and a deviation of $G$ from $G_0$ cannot influence the determination of $d_{\rm L}$.} }

\section{Measuring Newton's constant}

In this paper, the JLA compilation \cite{1401.4064} is adopted as an example. Covering a redshift range $0.01<z<1.3$, the JLA compilation assembles 740 SNIa samples. To study the evolution of model parameters with redshift, we employ a redshift tomographic method. To be specific, the JLA samples are binned into the following subgroups according to their redshifts: (1) $z<0.1$; (2) $0.1<z<0.2$; (3) $0.1<z<0.4$; (4) $0.1<z<1.3$; (5) $0.4<z<1.3$. In each subgroup, i.e. each redshift range, we assume the value of $G$, (i.e., $M_B^1$), is a constant. Also, we assume the relation Eq.~(\ref{eq11}) holds for each sample. However, the values of the nuisance parameters $\alpha$ and $\beta$ could be different for different subgroups. Therefore, if the values of $M_B^1$ at different redshift ranges are obtained, the difference of $G$ between different redshifts can be inferred.

The theoretical values of distance modulus $\mu$ strongly depend on the cosmological
parameters. To avoid model-dependence in our measurement of the redshift evolution of $G$, we need an alternative method to determine the value of $\mu$ at different redshifts. Future detectable GW events are expected to distribute in nearly the same redshift range as the SNIa data, and the $d_{\rm L}$ of GW events can be determined by the GW observations alone. For each SNIa sample with fixed redshift $z$, we can derive its distance $d_{\rm L}$ (or distance modulus $\mu$) from nearby GW events by a proper interpolation. When linear interpolation is used, the resulting $\mu$ and its error $\sigma_\mu$ at redshift $z$ can be calculated by
\begin{eqnarray}
\mu &=& \left[\frac{z_{i+1}-z}{z_{i+1}-z_{i}}\right]\mu_{i}+\left[\frac{z-z_{i}}{z_{i+1}-z_{i}}\right]\mu_{i+1},\label{eq22}\\
\sigma^2_{\mu} &=& \left[\frac{z_{i+1}-z}{z_{i+1}-z_{i}}\right]^2 \sigma_{\mu,i}^2+\left[\frac{z-z_{i}}{z_{i+1}-z_{i}}\right]^2 \sigma_{\mu,i+1}^2, \label{eq33}
\end{eqnarray}
in which $\mu_i$, $\mu_{i+1}$ are the distance moduli of the GW events, and $\sigma_{\mu,i}$, $\sigma_{\mu,i+1}$ their errors, at nearby redshifts $z_i$ and $z_{i+1}$, respectively.

\begin{figure}
\begin{center}
\centerline{\includegraphics[width=15cm]{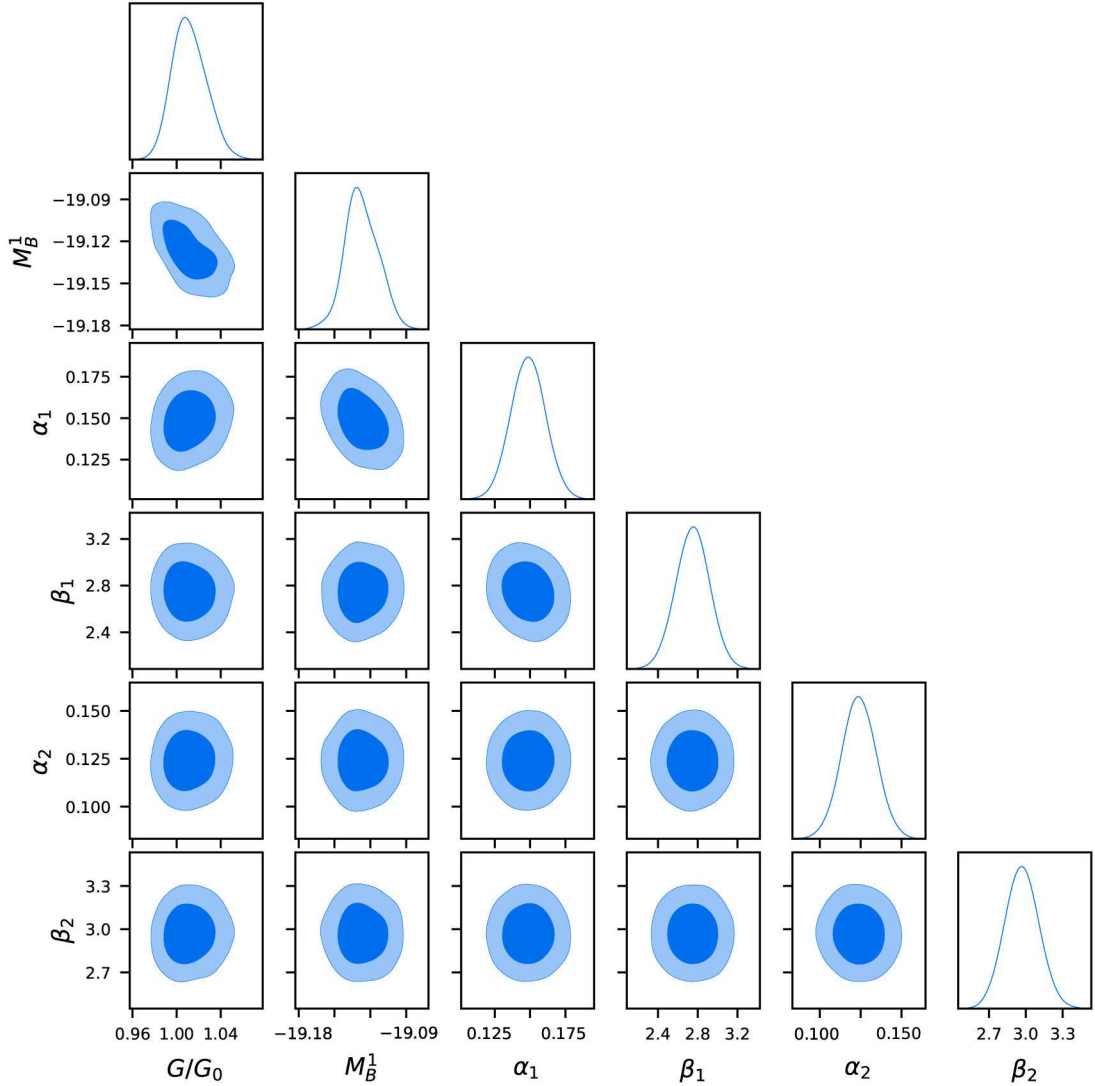}}
\end{center}\caption{Two-dimensional and one-dimensional constraint contours for six parameters, where we have considered the case of combining SNIa data at $z<0.1$ with those at $0.1<z<0.2$. The two-dimensional contours denote $1\sigma$ and $2\sigma$ constraints, respectively.}\label{f3}
\end{figure}

We first investigate the possible difference of
$G$ in the redshift ranges $z<0.1$ and $0.1<z<0.2$.
Considering the {SNIa} samples of these two subgroups, for each {SNIa}, $z$ is known, $\mu$ and $\sigma_{\mu}$ are derived from the interpolation 
of GW data, and the values of $m_{\rm B}^*$, $X_1$ and $C$ are given in \cite{1401.4064}. For these data in two bins, we have six parameters ($M_{\rm B}^1$, $\alpha_1$, $\beta_1$, $G/G_0$, $\alpha_2$, $\beta_2$), where ($\alpha_1$, $\beta_1$) and ($\alpha_2$, $\beta_2$) are the nuisance parameters in the first and second redshift bin, respectively, $M_{\rm B}^{1}$ is the {calibrated} absolute magnitude of{ SNIa} at $z<0.1$, and $G_0$, $G$ are the Newton's constant in the two redshift bins. Note that, throughout this paper we assume $G_0$, the value in the first $z$-bin, is equal to the $G$ value today, and that the calibrated absolute magnitude of SNIa in the second bin has been expressed by $M_{\rm B}^1$ and $G/G_0$. We apply the following $\chi^2$ calculation to obtain the constraints on six parameters,
\begin{eqnarray}\label{eq3}
{\chi}^2&=&\sum_{i}\frac{[\mu^{(i)}-(m_{\rm B}^*-M_{\rm B}+\alpha X_{1}-\beta{C})^{(i)}]^2}{{\sigma^2_{\mu^{(i)}}} +\sigma^2_{(m_{\rm B}^*-M_{\rm B}+\alpha X_{1}-\beta{C})^{(i)}}+\sigma^2_{s^{(i)}}}\nonumber \\ &+&
\sum_{j}\frac{[\mu^{(j)}-(m_{\rm B}^*-M_{\rm B}+\alpha X_{1}-\beta{C})^{(j)}]^2}{{\sigma^2_{\mu^{(j)}}} +\sigma^2_{(m_{\rm B}^*-M_{\rm B}+\alpha X_{1}-\beta{C})^{(j)}}+\sigma^2_{s^{(j)}}},
\end{eqnarray}
where $i$ and $j$ indicate the SNIa samples in first and second redshift bin respectively, and
$\sigma_s^2=({5\sigma_z}/{z\log10})^2+\sigma_{\rm lens}^2+\sigma_{\rm coh}^2$,
which accounts for the uncertainty in cosmological redshift due to peculiar velocity, the variation of magnitudes caused by gravitational lensing, and the intrinsic variation in SN magnitude not described by other terms \cite{1401.4064}.
We use $c\sigma_z=150{\rm km/s}$ and $\sigma_{\rm lens}=0.055z$ as suggested in \cite{Jonsson2010,1401.4064}. The values of $\sigma_{\rm coh}$ are adopted as in \cite{1401.4064}. Note that, in this calculation, we have ignored the weak correlation between different {SNIa} data, which only slightly changes the uncertainties of the constrained parameters.

Employing a modified CosmoMC package \cite{lewis}, we obtain the marginalised constraints on each parameter, which are listed in Table \ref{table1}, and the one-dimensional likelihood functions and two-dimensional contours are presented in Fig.~\ref{f3}. To measure the value of $G$ in different redshift ranges, we replace the second redshift bin $0.1<z<0.2$ with that in $0.1<z<0.4$, $0.1<z<1.3$, $0.4<z<1.3$, respectively. The corresponding constraints are also presented in Table \ref{table1}. We find that for each case, the uncertainty of $G/G_0$ is $\sim0.015$. These results show that, by combining the {SNIa} data and potential 3G GW data, the deviation of Newton's constant from $G_0$ at high redshifts can be expected to be constrained at $1.5\%$ level. The uncertainty in $G/G_0$ is mainly caused by the error bars of $\mu$ in the first redshift bin, which in turn are determined by the errors on $d_{\rm L}$ for GW events in the same redshift range. For comparison, we keep the second redshift bin as $0.1<z<0.2$, and change the first bin to $z<0.03$, $z<0.05$ and $0.05<z<0.1$. The corresponding uncertainties of $G/G_0$ become $0.018$, $0.017$, $0.022$ respectively, which are larger than $0.015$ as anticipated.

Note that the number of observable GW events, $N_{\rm GW}$, is quite uncertain. In order to test how the uncertainty of $G/G_0$ depends on $N_{\rm GW}$, we compare the following cases: (1) 1000 low-$z$ and 1000 high-$z$ GW events as above; (2) 300 low-$z$ and 1000 high-$z$ GW events; (3) 4500 low-$z$ and 1000 high-$z$ GW events; (4) 1000 low-$z$ and 2000 high-$z$ GW events. For each case, we consider the SNIa samples in two bins ($z<0.1$ and $0.1<z<0.2$), and derive the constraints of six parameters by a similar analysis as above. We find that the results are nearly the same in all cases, which is understandable: the
GW observations influence our results only through Eqs.~(\ref{eq22}) and (\ref{eq33}), and these two relations show that the values of $\mu$ and $\sigma_{\mu}$ for each SNIa depend only on its nearby GW events, and increasing or decreasing $N_{\rm GW}$ cannot significantly affect their values. Of course, if the redshift distribution of GW events is too sparse, i.e. $N_{\rm GW}$ is too small, the interpolation in Eqs.~(\ref{eq22}) and (\ref{eq33}) is 
not applicable any more, and the derived constraints on $G/G_0$ become unreliable. Therefore, to keep the stability of the results, the number of GW events should be comparable to, or even larger than, that of SNIa in the corresponding redshift ranges.

\begin{table}
\scriptsize
\caption{The uncertainties of six parameters for {SNIa} in different redshift bins combined with those in $z<0.1$.}
\label{table1}
\begin{center}
\begin{tabular}{|c| c c c c| }
    \hline
     & $0.1<z<0.2$ & $0.1<z<0.4$ &$0.1<z<1.3$ & $0.4<z<1.3$  \\
         \hline
    $G/G_0$ & $1.011\pm0.016$ & $1.010\pm0.015$ & $1.007\pm0.015$ & $0.997\pm0.016$ \\
    $M^1_{\rm B}$ & $-19.13\pm0.01$ &  $-19.13\pm0.01$ & $-19.13\pm0.01$ &  $-19.12\pm0.01$ \\
    $\alpha_1$ &$0.149\pm0.012$ & $0.149\pm0.012$ & $0.149\pm0.012$ & $0.149\pm0.012$ \\
    $\beta_1$ & $2.744\pm0.170$& $2.749\pm0.170$ & $2.747\pm0.169$ & $2.747\pm0.170$ \\
    $\alpha_2$ &$0.124\pm0.010$ & $0.140\pm0.008$ & $0.136\pm0.007$ & $0.119\pm0.016$ \\
    $\beta_2$ &$2.972\pm0.139$ & $3.037\pm0.100$ & $2.953\pm0.083$ & $2.738\pm0.161$ \\
         \hline

\end{tabular}
\end{center}
\end{table}

\section{Conclusions}

The {calibrated} intrinsic peak luminosity of a {SNIa} depends on the strength of gravity in the supernova's local environment. If $d_{\rm L}$ can be determined by independent observations, the {SNIa} can be treated as a tracker to measure the variation of gravitational constant $G$ in a wide redshift range. We propose to use the GW standard sirens distributed in {a} similar redshift range to determine $d_{\rm L}$ of {SNIa} by interpolation. As an application, we consider the recent JLA compilation of {SNIa} data, for which $d_{\rm L}$ values are assumed to be determined 3G GW observations. Splitting the {SNIa} samples into several subgroups according to their redshifts, we determine the value of $G$ in different redshift {ranges}. We find that the ratio $G/G_0$, where $G$ is the gravitational constant in the redshift $z>0.1$ and $G_0$ is that at $z_0<0.1$, can be determined at the level of $1.5\%$.

{As examples to compare our results with other constraints, we adopt $z=0.4$ ($z=0.9$) and assume a power-law cosmic time dependence, $G\propto t^{-\alpha}$, then the constraint $\Delta G(z)/G<0.015$ is equivalent to a constraint on the index of $|\alpha|\lesssim 0.04$ ($0.02$), which can be translated into $|({\rm d}G/{\rm d}t)/G|_{t=t_0}\lesssim 3\times 10^{-12}$ year$^{-1}$ ($1.5\times10^{-12}$ year$^{-1}$). This is of the same order as constraints from pulsars \cite{pulsar}, lunar laser ranging \cite{lunar} and BBN \cite{BBN} ($|({\rm d}G/{\rm d}t)/G|_{t=t_0}\lesssim 10^{-12}$ year$^{-1}$). Most importantly, the new method offers a novel and independent way to constrain Newton's constant $G$ over a wide redshift range $0<z<1.3$, which could also be extended to $0<z<2$ by future SNIa observations \cite{lssd-wfirst}.}

\begin{acknowledgments}
WZ is supported by NSFC Grants Nos.~11773028, 11653002 and 11633001. BSW is supported by a UK Science and Technology Facilities Council (STFC) research studentship.  BL is supported by the European Research Council (ERC-StG-716532-PUNCA) and the STFC through grant ST/P000541/1.
\end{acknowledgments}

\end{document}